\documentclass[showpacs,preprintnumbers,
amsmath,amssymb]{revtex4}

\begin{document}
\title{Superconductors with Superconducting Transition Temperatures Tc =
91K (1999), 120K (1994), 340K (2000), and 371K (1995):
Experimental Errors or a Technological Puzzle? Two-Component
Nonstoichiometric Compounds and the
Insulator--Superconductor--Metal Transition.}
\author{V. N. Bogomolov}
\affiliation{A. F. Ioffe Physical \& Technical Institute,\\
Russian Academy of Science,\\
194021 St. Petersburg, Russia}
\email{V.Bogomolov@mail.ioffe.rssi.ru}
\date{\today}
\begin{abstract}

  One of the reasons for the lack of understanding of  both the mechanisms
  underlying the HTSC phenomenon and of the instability of materials with $T_{c} > 300 \mathrm
  K$
  may be the widely accepted but wrong ideas about the types of chemical bonding in
  a substance and the radii of the atoms and ions. A revision of these concepts started in
  the beginning of the XX century in connection with the investigation of nonstoichiometric
  compounds (the berthollides) but did not reach a critical level until recently.
  Most of the HTSC materials, however, are actually two-component nonstoichiometric nanocomposites
  (the berthollides), whose components "dilute" or "stretch" one another. Each component
  resides in an "intermediate" state, which still remains poorly studied. For instance,
  in a system of particles having two paired electrons each (divalent particles),
  the unbroken electron pairs may start tunneling at a certain "medium" concentration
  (or in the temperature interval  ${T_{1} < T_{c}  < T_{2},)}$
  with the system becoming a Bose superconductor
  (the state between the insulator and the metal with BCS superconductivity). For univalent
  atoms (Na, Ag,…), however, such possibility realizes neither in the intermediate
  (the Mott transition) nor in the final state. Univalent metals are not superconductors.
  In the berthollides, however, a possible Jahn--Teller--Peierls-type instability may give rise
  to formation of diatomic molecules {\mathsurround=0pt(Na$_{2}$, Ag$_{2}$,…) with electron pairs, and superconductivity
  can set in. It is possibly such systems that were obtained by chance in experiments with
  univalent components and reported to have $T_{c}$  of up to $371 \mathrm K.$} Structures of a number of HTSC
  materials are considered.
  \end{abstract}
  \pacs{71.30.+h, 74.20.-z, 74.25.Jb}
  \maketitle
 In the recent decade, several reports from independent
sources announced synthesis of substances with the superconducting
transition occurring at $91 \rm K$\cite{1}, $120 \rm K$\cite{2},
$340 \rm K$\cite{3}, and $371 \rm K$\cite{4}. These experiments,
however, neither have received confirmation nor were continued. In
all the cases, the reason for this was the poor reproducibility of
the synthesis and the instability of the samples. It is this
feature, however, that is the main structural characteristic of
HTSC substances with the model of superconductivity used to
interpret the metallization of xenon under pressure \cite{5} and,
later, applied to two-component nonstoichiometric compounds with
incommensurate parameters of the constituents (matrix systems)
\cite{6}. In addition to the instability common to all the four
substances, they have also certain common physical features
characteristic of the given superconductor model, which permits
one to consider their appearance not as due to experimental errors
but rather as chance occurrences of technological nature
associated with structural instabilities.

Synthesis of superconductors with $T_{c} > 300 \rm K$ and
elucidation of the mechanisms underlying high-temperature
superconductivity are issues that are still awaiting solution. One
of the reasons for this may be the widely accepted but inadequate
concepts concerning the part played by the atomic and ionic states
(and by their radii) in the structure of a substance. A revision
of these concepts started in the beginning of the XX century in
connection with the investigation of nonstoichiometric compounds
(berthollides). The lack of understanding of their nature has not,
however, reached a critical level until recently. Indeed, we may
recall that most of the HTSC compounds are actually berthollides,
and their physical properties cannot be understood properly
without a correct description of their chemical structure. Rather
than being uniform chemical substances, the berthollides can be,
for instance, two-component nonstoichiometric nanocomposites, with
each of the components "diluting" or "stretching" the other. They
stabilize one another in an unusual, "intermediate" state, which
remains poorly studied. If, for instance, particles having two
paired electrons each (divalent particles) approach one another, a
real state may form within a narrow interval of concentrations (or
temperatures) between the insulator and the metal, a state where
pair tunneling has already started, and where, while their binding
energy has already begun to decrease, the electron pairs have not
yet split into fermions (as is the case with the final, metallic,
equilibrium state for $T
> T_{c}).$  The system transfers to the state of a Bose
condensate. A Bose superconductor forms between the insulator and
the metal with BCS superconductivity within a narrow and
structurally unstable region. For univalent atoms (Na, Ag,…),
however, such possibility realizes neither in the intermediate
(the Mott transition) nor in the final state. Univalent metals are
not superconductors. In the berthollides, however, a possible
Jahn--Teller--Peierls-type instability may give rise to formation
of diatomic molecules {\mathsurround=0pt(Na$_{2}$, Ag$_{2}$,…)}
with electron pairs and superconductivity can set in. It is
possibly such systems that were obtained by chance in experiments
with univalent components and reported to have $T_{c}$ of up to
$371 \rm K$. In the berthollides, stability is a consequence
primarily of their architecture rather than of chemical
interactions, a point that is frequently disregarded, thus
bringing about a distorted picture of physical phenomena. This may
come from a wrong determination of the type of chemical bonding in
the given structure because of the use of wrong values of the
radii and charges of the particles, a possibility substantiated by
a comparison of experimental data with quantum-mechanical
calculations of the radii of atoms and, particularly, of ions. For
instance, the radius of the {\mathsurround=0ptO$^{2-}$ ion is not
$1.4$\AA \:  but $0.6$\AA,\: and that of Cu$^{2+},$ $0.33$\AA \: rather
than $0.73$\AA \: \cite{7, 8}.}

   Two-component nonstoichiometric compounds are relatively inert
matrices with voids of diameter $D$ containing not ions but atoms,
whose diameters $d$ are slightly larger than those of the free
atoms but smaller than of atoms in a metal. When atoms are removed
from the metals (or when the number of the nearest neighbors
decreases), their diameter decreases by only $10\% -- 15\%$ (the
Goldschmidt contraction). Each component is a stoichiometric
substance, whose structure may be not necessarily stable in
itself. While atoms of metals, for instance, can have
close-to-metallic radii, the component itself may turn out to be
an unusual, "stretched" or "dilute" metal, and not a Fermi system
at all. Stoichiometric components are separated by a "gap" taking
up a sizable part of the nanocomposite volume \cite{9}.

 (1)    The case of $D \sim (3--10) d.$ Belonging here are, for
instance, the zeolites with substances embedded in the structural
voids \cite{10}. The clusters in the voids are domains of the
second phase. The system as a whole may be treated also as a
"supercrystal" \cite{11}.

 (2)    The case of $D \lesssim d.$ Atoms of the second
component (but not ions) stretch the matrix. This case, well known
in chemistry, relates directly to the nonstoichiometric compounds
(the berthollides or interstitial compounds). It can be
illustrated by the {\mathsurround=0pt $M\!e$B$_{6}$ system. A
sublattice of the metal atoms rather than of the matrix-doping
ions is inserted into the voids of the framework between the
B$_{6}$ octahedra. This is indicated by the framework stretching,
with {\it d}, the diameter of the atom, exceeding {\it D}, which
is the $M\!e$B$_{6}$ lattice parameter. For {\it M\!e} = Ce, La,
Ca, Sr, Ba the ratio $D/d$ is\ Ce $(4.14/3.26),$ La $(4.15/3.74),$
Ca $(4.15/3.95)$, Sr $(4.19/4.30),$ and Ba $(4.27/4.34).$ The
sublattices of Ce, La, and Ca are stretched, and those of Sr and
Ba, compressed, but at the same time the B$_{6}$} framework is
stretched. Metal ions could not stretch the framework. All these
compounds feature about the same melting points $T \sim 2200 \rm
K,$ which are determined by the rigid covalent {\mathsurround=0pt
B$_{6}$} framework \cite{12}. The B--B bond lengths of the five
nearest neighbors are $1.72$\AA.

(3) If $D \gtrsim d$ (perovskite-like structures), the atoms of
the second {\mathsurround=0pt$M\!e$} component inserted in the
voids of the matrix {\mathsurround=0pt$Mt$} can undergo a $3D$
Jahn--Teller--Peierls transition with formation of diatomic
molecules made up of atoms which occupy neighboring voids
\cite{6}. After all the voids have filled ({\mathsurround=0pt
$M\!e_{1}Mt_{1},$}) one can conceive of the following
possibilities:\\
(A) formation of regular structures of diatomic molecules;\\ (B)
formation of disordered structures.\\
In both cases, a matrix-stabilized second phase will appear. This
phase is at the same time "diluted" by the matrix.

     If not all of the voids are filled (a technological problem),
the composition of the substance can be written as
{\mathsurround=0pt $M\!e_{x}Mt_{1}$} for $x < 1,$ which likewise
suggests two possibilities:\\
(a) - domains with all voids filled will form (a version of cases
A and B);\\
(b) -   a component with a low concentration of diatomic molecules
forms.

     Realization of version (b) is a still more complex technological
problem, but it is such systems with very high $T_{c}$  that were
apparently obtained accidentally \cite{1,2,3,4}.

I.     If the inserted atoms are univalent, the diatomic molecules
will have two paired electrons and a binding energy $E_{m} =
E_{0}(D - d)/d$, where $E_{0}$ is the binding energy of the
{\mathsurround=0pt $M\!e$} metal \cite{6}. As the concentration
$N$ of such molecules increases, or the temperature decreases, the
thermal de Broglie wavelengths {\nolinebreak $\lambda_{dB} \sim
T^{-1/2}$}  of the electron pairs and the mean intermolecular
distances $N^{-1/3}$ may become comparable. It gives rise to
Bose-Einstein condensation at {\nolinebreak $\lambda_{dB}^{3}N >
2.612$ \cite{13}.} The condensation phenomenon of the
Bose-Einstein statistics is distorted, of course, by the presence
of molecular forses and by fact that it manifests itself not in
the gaseous state. But the same problems exist in the case of
liquid helium as well. The system will transfer from the insulator
to the Bose superconductor state with the transition temperature
${kT_{cn}=kT_{B}=3.31(h/2\pi)^{2}N^{2/3}/M=1.04
(h/2\pi)^{2}n^{2/3}/m,}$ where $n=2N$ and $m=M/2$ are the electron
concentration and mass, respectively. For $T_{B}=(15, \; 70,\;
325)$\!K, we obtain $n = (2\cdot10^{18},\; 2\cdot10^{19},\; 2\cdot
10^{20})\,\rm cm^{-3}.$ The transition temperature can, however,
be governed by the {\mathsurround=0pt$E_{m}$} energy as well, as
is the case with metals {\mathsurround=0pt($T_{cm}$).} As the
temperature increases to {\mathsurround=0pt $kT> kT_{cm} \sim \;
E_{m},$}\quad or the concentration rises to $n > n_{crit}$
(ultimately to $x = 1$), the bosons will separate into the
fermions, and the system will become Fermi type with the
temperature $T_{F} > T_{B}.$ Therefore, such systems can become
superconducting only within a narrow region of concentrations of
atoms (molecules) or temperatures ($T_{1} < T_{c} < T_{2}$). These
are either doped semiconductors, or "stretched" or "diluted"
metals; being actually real systems, they are not Fermi type but
rather unstable, and it is such systems that are dealt with here.
Such systems can become superconducting not only through formation
of electron pairs ($T_{cm})$ as is the case with metals, but as a
result of Bose condensation of the bosons present in the system
($T_{cn} = T_{B}$). An increase in the concentration of $M\!e$
atoms drives the insulator--superconductor--metal transition .

 II. If two paired electrons, rather than belonging to molecules,
are outer electrons of a divalent atom, for instance, of Mg, the
components need not necessarily be incommensurate. One has only to
produce in the voids of an inert matrix a local concentration of
the metal atoms $M\!e_{x}$ within a certain interval $x < 1$
(provided there are no domains with $x = 1$, as is the case with
an equilibrium Fermi metal). The mechanism of preservation of
electron pairs in metals, if they form for $T < T_{cm}$ from the
outset as superconductors, is described by BCS theory (the Cooper
pairs). Univalent metals do not have such "relic" states, and this
is what accounts for their not exhibiting superconductivity..
Compression of a metal usually reduces {\mathsurround=0pt$T_{c}.$}
Hence, Cooper pairs in a "stretched" metal are bound stronger.
They are much closer to the state of atomic electron pairs.

  {\mathsurround=0ptMgB$_{2}$.  \quad   It is a version of such a "diluted metal" that possibly
  is realized in the MgB$_{2}$ system. The divalent magnesium is "diluted"
  by boron layers in the form of a covalently bonded hexagonal network.
  The MgB$_{2}$} lattice parameters are $c = 3.40$\AA,\; $a = 3.05$\AA  \cite{14}.
  Each B atom has three nearest neighbors. The B--B distance, $1.77$\AA, is larger
  than the dimensions of the atom, $1.55$\AA  \cite{8}, than that in the
  {\mathsurround=0ptB$_{2}$}
  molecule $(1.59$\AA), and that in {\mathsurround=0pt$M\!e$B$_{6}$}\; $(1.72$\AA), which has five nearest
  neighbors. Having a nearly covalent diameter $(3.08$\AA), the magnesium atoms
  stretch the boron atom network. This produces conditions favorable for the formation
  of {\mathsurround=0ptB$_{2}$} molecules. The Mg atoms confined between the B layers are also slightly
  stretched along the c axis as compared to their diameters in the metal $(3.20$\AA).
  This system revealed two energy gaps proportional to {\mathsurround=0pt$T_{cm1}=15$K and
  $T_{cm2}=45$K.}

  {\mathsurround=0pt$M\!e_{3}$C$_{60}.$\quad    Another illustration of superconductors with a stretched component
  (version $D \lesssim d$) is provided by the $M\!e_{3}$C$_{60}$ fullerides.
  Their $T_{c} = T_{cm} \sim E_{0}(D - D_{0})/D_{0},$ where $E_{0}$ is the binding energy
  of the C$_{60}$ spheres, $D_{0}$ is the fullerene lattice parameter, and $D$ is this
  parameter increased by incorporation of $M\!e_{3}$ clusters into the octahedral voids \cite{6}. $T_{cmax}\sim
  40$K.}

  {\mathsurround=0ptNa$_{x}$CoO$_{2}-1.3$H$_{2}$O.\quad     The superconductivity $(T_{c}\sim 4$K)  observed in the
  Na$_{x}$CoO$_{2}-1.3$H$_{2}$O  system is due to the univalent sodium, possibly as
  a result of formation of weakly bound Na$_{2}$} molecules. The optimum concentration
  of Na is $x = 0.3$  \cite{15}, i.e., it is "intermediate", as in many other cases. It is stabilized by the {\mathsurround=0pt H$_{2}$O} molecules ( an "additional diluent"  or third component ).

   {\mathsurround=0ptNa$_{x}$-NH$_{3}$.\quad     The same molecules (Na$_{2}$) formed undoubtedly in the
   Na$_{x}$-NH$_{3}$}
   solutions as well $(T_{c}\sim 200$K) \cite{16}. The instability of this system originates from fast
   phase separation in the liquid, and the decrease in the solution density can be traced
   to the increase of the atomic volume of sodium in diatomic molecules and the interphase
   "gap" \cite{6,9} rather than to the formation of "bubbles" \cite{16}.

  {\mathsurround=0ptY$_{2}$Ba$_{4}$Cu$_{6}$O$_{14}$.\qquad  The well studied system
  Y$_{2}$Ba$_{4}$Cu$_{6}$O$_{14}$}
  $(85 \rm K)$\cite{17} can also be treated in terms of the above scheme. Its lattice
  parameters are $c=11.681$\AA$=3\cdot(3.894)$\AA,\;\qquad $a=3.886$\AA,\qquad
  $b=3.827$\AA \cite{17}.
  This structure   may be considered also as a slightly distorted simple cubic lattice
  of spheres with an average diameter of $3.86$\AA. \; It corresponds to the average diameter
  of the Y and Ba atoms in these spheres of $3.87$\AA. \; The diameters of the free atoms are
  $3.38$\AA \; and $4.12$\AA, \; and those of the ions {\mathsurround=0ptY$^{3+}$ and Ba$^{2+},$} $1.28$\AA \; and
  $1.73$\AA.\;
  The diameters of the atoms in the metal are $3.56$\AA \; and $4.34$\AA. The density of this
  lattice is $0.52$\; ("dilute metal"). The remainder of the volume, $0.48$, \;is taken up by
  the Cu--O octahedra and pyramids and the interphase "gap" \cite{9}. Each oxygen atom
  is also bonded covalently to six or eight nearest oxygen atoms. Sulfur, an analog of oxygen,
  also exhibits a tendency to formation of structures of a molecular type {\mathsurround=0pt(S$_{2}$, S$_{6}$,
  S$_{8}$). In the case of oxygen, this process requires a "seed" (Cu) for its realization.
  The Cu$_{3}$O$_{7}$} component is a stoichiometric compound with O--O covalent bonding
  \cite{18}. It is stable naturally only in the framework of the {\mathsurround=0ptY-Ba$_{2}$ lattice.
  Both components stabilize one another. The Y and Ba sublattices are "diluted", and
  the Y sublattice, "stretched" as well (Y$_{2}$). Hence, the volume of the
  Cu$_{3}$O$_{7}$}
  component is not governed by the ionic diameters, either traditionally accepted $(1.46$\AA \;
  and $2.8$\AA) \;or calculated $(0.65$\AA \;and $1.20$\AA) \cite{7, 8}.

  {\mathsurround=0ptNa$_{0.05}$WO$_{3}.$    The properties of two-component
nonstoichiometric compounds
 become most clearly pronounced in the $M\!e_{x}$WO$_{3}$ tungsten bronzes \cite{6}.
 The bronze with an average composition Na$_{0.05}$WO$_{3}$} was found to be superconducting
 with $T_{c} = 91 \rm K$ and to have an energy gap $E \sim 160 \rm K$ \cite{1}. The substance is
 unstable. The parameter $D = 3.78$\AA. In the Na metal, $d = 3.72$\AA. In this case,
 for $E_{0} = 1$eV in sodium we obtain for the gap energy $E_{m} = E_{0}(D - d)/d = 0.016$eV,
 which implies that the transition is energy driven. Free {\mathsurround=0ptNa$_{2}$ molecules have a binding
 energy of $0.35$ eV. In the WO$_{3}$} matrix, their energy is $20$ times smaller $(0.35/0.016)$.
 Their detection is made difficult by the fact that the modulation of the period is only $0.016$.
 {\mathsurround=0ptWO$_{3}$ is a standard stoichiometric compound (W$^{6+}$ - O$^{2-}).$} However, in this
 compound each oxygen atom has likewise eight nearest oxygen atoms. Therefore, its
 physico-chemical bonding is also a quantum-mechanical superposition of several types
 of states \cite{18}. Remarkably, replacement of Na by Rb or Cs, for which $D < d$,
 lowers sharply $T\!_{c}$ (from $91 \rm K$ down to $2--3 \rm K$). Substitution of Na by Li or Ag
 $(d \sim 3.10$\AA) was not tried, because Na, Rb, and Cs were considered as dopants only
 (i.e., in the ionic form), rather than a second component in a nonstoichiometric compound \cite{1}.

 Ag--Y--Ba--Cu--O.      Substitution of this type was made earlier in the Y--Ba--Cu--O system.
 One doped it with silver, which raised $T_{c}$  to $120 \rm K$\cite{2}.

   {\mathsurround=0ptAg$_{x}$Pb$_{6}$CO$_{9}$.   The use of silver in the Ag$_{x}$Pb$_{6}$CO$_{9}$ system
 also brought about an increase in $T_{c}$ to $340 \rm K$\cite{3}. The "colossal electrical
 conductivity $>10^{9}$ Ohm$^{-1}$cm$^{-1}$" observed to occur in the Ag$_{5}$Pb$_{2}$O$_{6}$
 system between $210 \rm K$ and $525 \rm K$ is assigned \cite{19} to a heating-induced segregation of the
 Ag component in channels of the Ag$_{3}$Pb$_{2}$O$_{6}$ lattice. The composition proposed
 for this compound in \cite{19} is Ag$_{2}$[Ag$_{3}$Pb$_{2}$O$_{6}].$} If this state is indeed
 superconductivity, it becomes realized within the interval ($T_{1} < T_{c} < T_{2}$)
 (i.e., under heating).

   {\mathsurround=0ptYBa$_{2}$Cu$_{3}$Se$_{7}$.  $ T_{c}$ can be increased not only by reducing $d$ but by
   increasing $D$ as well. The oxygen in the standard Y--Ba--Cu--O system was replaced
   by selenium, which brought about an increase in the lattice parameter, after which
   YBa$_{2}$Cu$_{2}$Se$_{7}$} exhibited $T_{c} \sim 371 \rm K$\cite{4}.

     One can conceive also of a specific class of two-component  nanocomposites of
     a "dynamic" type. These are condensates of molecular, or noble gases. In these
     condensates, a certain number of virtual excimer molecules, for instance, {\mathsurround=0ptXe$_{2}$* \cite{20}
     with two paired electrons, are present in the lattice of Xe atoms residing in the ground
     state ("matrix"). It is supposed, that increasing the Xe$_{2}$* concentration either by
     applying pressure or through the action of catalysts (atoms of metals) could give rise
     to Bose superconductivity with high $T_{c}$ \cite{5}. The investigation of the Mott
     transition in metal--inert-gas nanocomposites and of adsorption forces (sorption compounds)
     is dealt with in a large number of publications (see, e.g., (\cite{21, 22, 23, 24}).
     It is quite possible that a W-Ar sorption compound forms in the gas sheath surrounding
     the tungsten filament in the course of metal vaporization in conventional gas-filled tubes.
     Metallization of this layer brings about a drop in the voltage across the filament with
     increasing current \cite{25}. One cannot exclude here, however, the possibility
     that the increase in the conductivity of the gas is actually the result of its ionization.

     Thus, the major technological difficulty in the way of developing HTSC materials
     consists in preparing stable systems of particles with paired electrons (for instance,
     divalent atoms or molecules) in concentrations at which the electron pairs already are
     capable of tunneling (Bose condensation) while not yet separating into single electrons,
     either as a result of conventional chemical interactions or through a decrease in their
     binding energy in an effective dielectric medium. Such an intermediate state corresponds
     to a "stretched" or "diluted" substance, which can be stabilized by using "solid
     solvents" as matrices. The highest $T_{c}$ were obtained with univalent metals (not
     superconductors), because their diatomic molecules can exist in the matrices in
     low enough concentrations. In some cases such nonequilibrium concentrations are
     possibly "frozen" by chance fluctuations in the technology of synthesis
     (\cite{1, 2, 3, 4}).  Stability has thus far been reached apparently at the expense
     of a low $T_{c}.$} In searching for efficient methods of structural stabilization,(with help of some amount of inert particles as "additional diluent" (third component), for instance)
     a better understanding of the quantum-mechanical interactions in the structure of
     each component is of crucial importance. Such methods of stabilization can be found
     only by using realistic atomic and ionic radii rather than conventional quantities,
     which are nothing else but a consequence of postulated bonding types. The matrices
     in such nanocomposites also exist in an unusual state because of the unusual structure
     and contact interaction with the second component. This complicates greatly the
     investigation and description of such systems.


\begin{references}
{\mathsurround=0pt
\bibitem{1}   Shengelaya, A., Reich, S., Tsabba, Y., and  Muller, K.A.
Electron spin resonance and magnetic susceptibility suggest
superconductivity in Na doped WO$_{3}$  samples.  {\it Eur. Phys.
J.} {\bf B12}, 13-15   (1999).
\bibitem{2}   Yanmaz, E., Multu, I.H., Kucukomeroglu, T.,
Altunbas, M.     Ag-doped 120K YBa$_{2}$Cu$_{3}$O$_{7-d}$
superconductors prepared by the flame-quench-melt-growth (FQMG)
method {\it Supercond. Sci. Technol.}   {\bf 7},  903 - 907
(1994).
\bibitem{3}    Djurek, D., Medunic, Z., Tonejc, A., Paljevic, M.
superconductivity from Pb$_{3}$CO$_{5}$ - Ag$_{2}$O (PACO) system
{\it Physica C} {\bf 341-348},   723-725    (2000).
\bibitem{4}   Shabetnik,V.D., Butuzov, S.Yu.,
Plaksii, V.I.  High-temperature superconducting compound
YBa$_{2}$Cu$_{3}$Se$_{7}$ with $T_{c} =371 \rm K$.      {\it
Techn. Phys. Lett.} {\bf 21}, 382 - 384   (1995).
\bibitem{5}   Bogomolov, V.N.   Metallization of molecular
condensates and superconductivity.             {\it Techn.  Phys.
Lett.} {\bf 28},  211-215  (2002).
\bibitem{6}   Bogomolov, V.N.  Superconductivity of
the two-component non-stoichiometric compounds with incomensurate
sublattices.   {\it e-print,  cond-mat/0304561.}
\bibitem{7}    Slater, J.C.
Atomic radii in crystals. {\it J. Chem. Phys.} {\bf 41}, 3199-3204
(1964).
\bibitem{8}    Waber, J.T., Cromer, Don T.    Orbital radii of
atoms and ions.    {\it J. Chem. Phys.}  {\bf 42},  4116-4123
(1965).
\bibitem{9}  Bogomolov, V.N.    Capillarity phenomena in extremely thin zeolite
channels and metal-dielectric interaction.      {\it Phys. Rev.}
{\bf 51}, 17040 - 17045  (1995).
\bibitem{10}   Bogomolov,V.N.       Liquids in
ultrathin channels (thread and cluster crystals). {\it Usp. Fiz.
Nauk} {\bf 124},  171 - 182     (1978).   [{\it Sov.Phys. Usp.}
{\bf 21}, 77 - 86 (1978)].
\bibitem{11}  Arita, R., Miyake, T., Kotan, T.,
Schilfgaarde, M.van., Oka, T., Kuroki, K., Nozue, Y., Aoki, H.
Electronic properties of alcali-metal loaded
zeolites-"supercrystal" Mott insulator. {\it e-print,
cond-mat/0304322}.
\bibitem{12}  Mandelcorn,L. (ed)  in {\it Non-Stoichiometric Compounds}  (Academic Press, N.Y. and London, 1964)..
\bibitem{13} London, F.    On the Bose-Einstein condensation.
Phys.Rev.,
{\bf 54}, 947-954  (1938).
\bibitem{14}   Xu, M., Kitazawa, H., Takano, Y., Ye,
J., Nishida, K., Abe, H., Matsishita, A., Kido, G.    Single
crystal MgB$_{2}$ with anisotropic superconducting properties.
{\it e-print, cond-mat/0105271.}
\bibitem{15}    Schaak, R.E.,  Klimczuk, T.,
Fool, M.L., Cava, R.J.     Superconductivity phase diagram of
Na$_{x}$CoO$_{2}$ - 1.3 H$_{2}$O .  {\it Nature} {\bf 424},
527-529 (2003).
\bibitem{16}  Ogg, R.A.Jr.   Bose-Einstein condensation of trapped electron pairs.
Phase separation and superconductivity of metal-ammonia solution.
{\it Phys. Rev.}   {\bf 69},  243-244  (1946).
\bibitem{17}  Izumi, F., Asano, H.,
Ishigaki, T., Ono, A., Okamura, P.   Crystal structure of
Ba-Y-Cu-O superconductor as revealed by Rietveld analysis of x-ray
powder diffraction data.  {\it Jap. J. Appl. Phys.} {\bf 26},
L611-L612 (1987).
\bibitem{18}  Krebs, H. {\it Grundzuge Der
Anorganischen Kristallchemie}  (ed  Enke, F.,  Stuttgart, 1968).
\bibitem{19}   Djurek, D., Medunic, Z., Paljevic, M., Tonejc, A.
Colossal electric conductivity in Ag-defect
Ag$_{5}$Pb$_{2}$O$_{6}$. {\it e-print, cond-mat/0310011.}
\bibitem{20}   Tilton, R.A., Flynn, C.P. Sharp
optical spectra of impurities in metal.   {\it Phys. Rev. Lett.}
{\bf 34}, 20-23  (1975).
\bibitem{21}  Phelps, D.J., Avci, R., Flinn, C.P.      Metal-insulator
transition in metal-rare gas-alloys. {\it Phys. Rev. Lett.,} {\bf
34}, 23-26  (1975).
\bibitem{22}   Endo, H., Eatah, A.I., Wright, J.G., Cusack, N.E.    A
metal-nonmetal transition in argon-copper system.  {\it J. Phys.
Soc. Of Japan,} {\bf 34},  666-671  (1973).
\bibitem{23}    Leung, A.W.K., Kaup, J.G.,
Bellert, D., McGaffrey, J.G., Breckenridge, W.H.  Spectroscopic
characterization of excited Ca$(4s4d\delta3D_{J})$
RG$(^{3}\Delta_{1,2})$ states (RG=Ar, Kr, Xe): no "heavy-atom"
mixing of RG$(nd\delta)$ character into the wave functions. {\it
J. Chem. Phys.,} {\bf 111}, 981-987  (1999).
\bibitem{24} Burda, J.V., Runeberg, N., Pyykko, P.    Chemical bond between
noble metals and noble gases. Ab initio study of neutral diatomic
NiXe, PdXe and PtXe.    {\it Chem. Phys. Lett.,} {\bf 288},
635-641 (1998).
\bibitem{25}  Markov, G.A., Melnikova, N.V., Khon, Yu.
Anomalously high conductivity of deformed metals at elevated
temperatures. {\it Techn.  Phys. Lett.,} {\bf 22},  680-681
(1996).}

\end{references}
\end{document}